\documentclass[graybox]{svmult}

\usepackage{mathptmx}       
\usepackage{helvet}         
\usepackage{courier}        
\usepackage{type1cm}        
\usepackage{makeidx}         
\usepackage{graphicx}        
\usepackage{multicol}        
\usepackage[bottom]{footmisc}

\makeindex             

\begin{document}

\title*{Gamma-rays from millisecond pulsars in Globular Clusters}
\author{W\l odek Bednarek}
\institute{Department of Astrophysics, University of \L \'od\'z, ul. Pomorska 149/153, 90-236 \L \'od\'z, \email{bednar@uni.lodz.pl}}
%
%
\maketitle

\abstract*{Globular clusters (GCs) with their ages of the order of several billion years contain many final products of evolution of stars such as: neutron stars, white dwarfs and probably also black holes. These compact objects can be at present responsible for the acceleration of particles to relativistic energies. 
Therefore, $\gamma$-ray emission is expected from GCs as a result of radiation processes occurring either in the inner magnetosperes of millisecond pulsars or in the vicinity of accreting neutron stars and white dwarfs or as a result of interaction of particles 
leaving the compact objects with the strong radiation field within the GC.
Recently, GeV $\gamma$-ray emission has been detected from several GCs by the new satellite observatory ${\it Fermi}$. Also Cherenkov telescopes reported interesting upper limits
at the TeV energies which start to constrain the content of GCs.
We review the results of these $\gamma$-ray observations in the context of recent scenarios for their origin.}

\abstract{Globular clusters (GCs) with their ages of the order of several billion years contain many final products of evolution of stars such as: neutron stars, white dwarfs and probably also black holes. These compact objects can be at present responsible for the acceleration of particles to relativistic energies. 
Therefore, $\gamma$-ray emission is expected from GCs as a result of radiation processes occurring either in the inner magnetosperes of millisecond pulsars or in the vicinity of accreting neutron stars and white dwarfs or as a result of interaction of particles 
leaving the compact objects with the strong radiation field within the GC.
Recently, GeV $\gamma$-ray emission has been detected from several GCs by the new satellite observatory ${\it Fermi}$. Also Cherenkov telescopes reported interesting upper limits
at the TeV energies which start to constrain the content of GCs.
We review the results of these $\gamma$-ray observations in the context of recent scenarios for their origin.}

\section{Introduction}

Globular Clusters are huge concentrations of old stars with masses of the order of the Solar mass or lower. These $10^5-10^6$ stars are contained within a spherical volume of a few parsec.
About $\sim 150$ of the known clusters have been observed creating a spherical halo around the Galaxy. They are at typical distances of $\sim 10$ kpc from the Sun (e.g. Harris~1996). GCs also contain remnants of evolution of stars with masses $M_*>1M_\odot$, which can be responsible for acceleration of particles to high energies. In fact, more than a hundred of millisecond pulsars (MSPs) have been discovered within globular clusters (e.g. Camilo \& Rasio~2005) and large fraction of objects belonging to GCs are X-ray emitters. These sources are identified with the Cataclysmic Variables (accreting White Dwarfs) or Low Mass X-ray Binaries  (LMXBs), i.e. accreting neutron stars in the binary systems. Due to the presence of these compact objects, GCs have been suspected to be sources of non-thermal processes which can turn to the production of high energy radiation. However, only recently GCs have been detected in high energy $\gamma$-rays (HE: E$>$100 MeV) by the ${\rm Fermi}$-LAT telescope (Abdo et al.~2010a). In this paper we review the state of the knowledge on the observations and modelling of these oldest structures in the Galaxy.

\section{The stellar content of Globular Clusters}

A large number of solar type stars in a small volume create a very well defined background radiation field (the mass to luminosity ratio in GCs is close to $\sim 2$, e.g. van den Bosch et al.~2006). The observed luminosity of the globular clusters, $L_{\rm GC}$, and the density profile for the distribution of the stars inside it (Michie 1963a,b,c,d), 
\begin{eqnarray}
D(R) = \left\{ \begin {array}{ll}
1,                                                 & R < R_{\rm c} \\
(R_{\rm c}/R)^2,                         & R_{\rm c} < R < R_{\rm h} \\
(R_{\rm c}R_{\rm h})^2/R^4, & R_{\rm h} < R < R_{\rm t}, \end{array} \right.
\label{eq1}
\end{eqnarray}
\noindent
allows us to calculate the energy density of stellar photons inside the cluster (Bednarek \& Sitarek~2007),
\begin{eqnarray}
U_{\rm rad} = {{L_{\rm GC}}\over{c R_{\rm t}^2}} 
{{18R_{\rm t}^2 - 3\sqrt{6R_{\rm t}^3R_{\rm c}} -2R_{\rm c}^2}
\over{6(\sqrt{6R_{\rm t}R_{\rm c}^3} - 2R_{\rm c}^2}},
\label{eq2}
\end{eqnarray}
\noindent
where $R_{\rm c}$ is the core radius, $R_{\rm t}$ is the tidal radius, and the half mass radius is related to those both by $R_{\rm h} = \sqrt{2R_{\rm c}R_{\rm t}/3}$.
For typical parameters, $R_{\rm c} = 0.5$ pc, $R_{\rm t} = 50$ pc, and $L_{\rm GC} = 10^5$ L$_\odot = 3\times 10^{38}$ erg s$^{-1}$, we obtain $U_{\rm rad}\approx 300$ eV cm$^{-3}$.
The energy density of this stellar radiation field clearly dominates over the energy density of the Cosmic Microwave background Radiation (CMBR), $U_{\rm CMBR} = 0.25$ eV cm$^{-3}$. However, the photon number densities of the stellar and CMBR are comparable.

\begin{figure}[t]
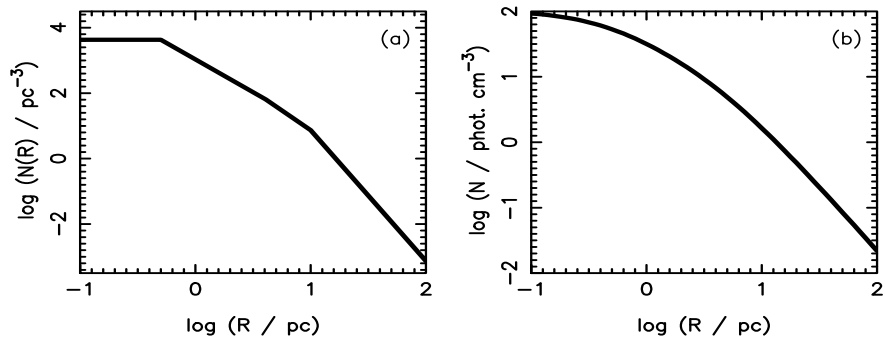

\vskip 4.7truecm
\includegraphics{bednarf1a.eps}
\includegraphics{bednarf1b.eps}
\caption{The density of stars (a) and density of stellar photons (b) as a function of distance $R$ from the center of a typical globular cluster with the mass $10^5$ M$_\odot$, the core radius $R_{\rm c} = 0.5$ pc,  the half mass radius $R_{\rm h} = 4$ pc, and the tidal radius $R_{\rm t} = 50$ pc (from Bednarek \& Sitarek~2007).}
\label{fig1}
\end{figure}

The radiation field inside GC can be defined as a function of distance to its center by normalizing the density profile of the distribution of stars inside GC to the total number of stars contained in the GC such as $N_\star^{\rm tot} = 10^5$ M$_\odot$.  
The number of stars as a function of the distance is then $N_\star(R) = A D(R) dR$ (where $A = N_\star^{\rm tot}/[2R_{\rm c}-2R_{\rm c}^2/(3R_{\rm h})-
R_{\rm c}^2R_{\rm h}^2/(3R_{\rm t}^3)]\approx N_\star^{\rm tot}/(2R_{\rm c})$). As a first approximation can be assumed that in average stars inside GC has parameters similar to the Sun, i.e. the surface temperature $6000$ K and the radius $7\times 10^{10}$ cm. The density of stellar photons at a specific distance from the center of GC is calculated by integration over the whole distribution of stars inside the GC. These density profiles for stars and stellar radiation inside the GC are shown in Fig.~\ref{fig1}. 
Thus, GCs belongs to the rare class of sources in which the soft radiation field within the cluster is very well defined. This is very important since it limits the number of free parameters when modelling
high energy radiation in these objects.

\section{Compact objects within Globular Clusters}

Due to their ages,  GCs contain large population of compact objects which are the final products of the evolution of stars with masses $>1M_\odot$. The white dwarfs (Cataclysmic variables) and neutron stars (LMXBs and MSPs) has been detected in many globular clusters.
In contrary to isolated compact objects in the Galaxy, the evolution of compact objects
within the GCs is strongly influenced by their frequent encounters with normal stars. They turn to the formation of compact binary systems. Therefore, a number of the X-ray sources related to such binary systems has been observed (see below). Here we briefly discuss these compact binaries (or their remnants) within the GCs. 

\subsection{Millisecond pulsars}

About 140 millisecond pulsars (MSPs) have been discovered in radio observations in 26 GCs
(for summary see Freire~2009). They are believed to be the result of spun-up to millisecond periods 
of old neutron stars which accumulate the angular momentum with the matter accreting from the companion star (Alpar et al.~1982). Some of these MSPs are isolated at present due
to complete evaporation-destruction of the companion star.
The largest population of MSPs has been detected in 47 Tuc (23 MSPs) and Ter 5 (33 MSPs).
However, many more are expected based on simulations of their formation 
in stellar encounters which should occur very frequently in GCs. For instance Ter 5 may contain of the order of $\sim 200$ MSPs (Fruchter \& Goss~2000, Kong et al.~2010). In fact, there should exist a link between stellar encounter rate in  GC
and the number of MSPs and LMXBs (Pooley 2003, Abdo et al.~2010).
The stellar encounter rate is estimated based in the known core stellar density and core radius for each particular GC (Verbunt \& Hut~1987). In this way it is possible to predict the MSP population.
It has been recently argued (Hui et al.~2010) that the population of MSPs in GCs have 
different X-ray emission properties from the MSPs in the Galactic field
(note however Becker et al.~2010 for counter opinions).
It is not clear whether proprieties of MSPs observed at other photon energies (e.g. in $\gamma$-rays) do not differ significantly as well. 

\subsection{Cataclysmic Variables and LMXBs}

First X-ray observations of GCs (${\it Uhuru}$ and ${\it OSO-7}$ satellites) reported a relatively large number of luminous low mass X-ray binaries (LMXBs) per unit mass within GCs with respect to those observed in the Galactic field (Clark~1975, Katz~1975). 
The observed emission was variable with power as high as $L_x > 10^{35}$ erg s$^{-1}$.  More sensitive instruments a population of low-luminosity X-ray sources 
with luminosities $L_x\sim 10^{31}-10^{33}$ erg s$^{-1}$ within  GCs has been reported by the more sensitive instruments such as ${\it Einstein}$ and ${\it ROSAT}$ (Hertz \& Grindlay ~1983, Verbunt~2001). However, only recently a firm identification has been possible thanks to deep observations with the Chandra telescope, which was able to resolve $\sim$1400 X-rays sources.
Many of these sources have been identified with the LMXBs and Cataclysmic Variables. The number of LMXBs within specific GC shows clear correlation with the encounter rate (Pooley et al.~2003).
Similar correlation has been also discovered between the number of Cataclysmic Variables within specific GC and the encounter rate (Pooley \& Hut~2006).
Recently, Pooley~(2010) proposed that the correlation between the number of X-ray sources within specific GC and the encounter rate has different slope for the normal and core collapsed GCs. Such effect is consistent with predictions of simulations of different types of GCs (see Fregeau~2008).
In conclusion, not only the amount of MSPs within the GCs are correlated with the encounter rates but also the amount of LMXBs and Cataclysmic Variables show such correlations.

\subsection{Intermediate mass black holes ?}

It has been suspected on theoretical grounds that in the central regions of the core collapsed GCs intermediate mass black holes (IMBHs, $W_{\rm BH}\sim 10^3 M_\odot$) should exist. Observational arguments for the existence of a black hole with mass $\sim 3\times 10^3M_\odot$ in M~15 has been reported by Gerssen et al.~(2002). On the other hand, observations of radio emission from a supposed accreting IMBHs in Tuc 47 and NGC 6397 put upper limits on the black hole mass equal to a few hundred $M_\odot$ (De Rijcke et al.~2006). Similar upper limits of same order 
of magnitude have been reported on the masses of IMBHs in Galactic GCs. Note that only recently, the evidence of the existence of the intermediate black hole ($\sim 500M\odot$) inside another galaxy has been reported by Farrell et al.~(2009). IMBHs in GCs can not certainly be strong persistent emitters since such sources are not observed in the Galactic GCs. However, transient activity, e.g. as a result of capturing-disruption of a star from the globular cluster, may happen rarely and can not be excluded at present.

\section{Non-thermal emission from Globular Clusters}

Here we review the results of observations of GCs which indicate the existence of relativistic particles. The presence of such particles is explicitly proved by detection of GeV $\gamma$-ray emission from GCs. The existence of relativistic electrons of diffusive nature within the globular cluster (or near the compact objects
inside GCs) can be also  envisaged  through the detection of non-thermal lower energy radiation, e.g. in the form of diffusive radio or X-ray emission. 

\subsection {Low energy radiation}

First reports on the extended non-thermal X-ray emission appeared in the results from the  Einstein and ROSAT observatories (e.g. Hartwick et al.~1982, Krockenberger \& Grindlay~1995). This emission has been identified in many cases as a result of the acceleration of particles of heating of gas in the interaction of GCs with the gas in Galactic halo. More recently, extended X-ray emission  has also been reported from the direction of 6 GCs using  ${\it Chandra}$ data (Okada et al.~2007). However, most of these X-ray sources appeared not to be related with the GCs but rather with the background clusters of galaxies. Other analyses of ${\it Chandra}$ data such the one reported by Hui et al.~(2009) show X-ray emission from 10 GCs and identifies most of this emission with compact sources. Recently, Eger et al.~(2010) observes diffuse X-ray emission from Ter 5 well described by a hard power spectrum.

We conclude that up to now there are not clear reports on the discovery of non-thermal diffusive X-ray emission from GCs. It looks that most of the reports can be interpreted as
the X-ray emission from the compact sources within the GCs or the X-ray emission from the background clusters of galaxies.

\subsection{Gamma-rays}

High energy $\gamma$-ray emission (HE: E$>$100 MeV, VHE: E$>$100 GeV) from a population of GCs has been discovered only recently. The prediction of very high energy $\gamma$-ray emission from GCs is on the level which gives a realistic chance for detection by the next generation of Cherenkov telescopes (HESS II, MAGIC II). The production of VHE $\gamma$-rays by leptons accelerated in the vicinity of MSPs seems to be quite certain. However, the level of this emission is difficult to estimate based on the present stage of knowledge since it depends on the number of MSPs and the diffusion of relativistic particles within the cluster. Its discovery may have to wait for the construction the next generation of instruments such as CTA.

\subsubsection{GeV emission}

The first upper limits on the GeV $\gamma$-ray emission were derived based on the 
EGRET and COMPTEL observations on the board of Compton GRO (Michelson et al.~1994, Manandhar et al.~1996, O'Flaherty et al.~1995). First clear detections of two globular clusters, Tuc 47 and Ter 5, were claimed based on the data collected by the ${\it Fermi}$-LAT detector (Abdo et al.~2009a, Kong et al.~2010). The analysis of 13 GCs allowed to associate 8 $\gamma$-ray sources in the directions of known GCs (see Table I in Abdo et al.~2010a). The $\gamma$-ray spectra of these GCs are incredibly similar (see Fig.~2 in Abdo et al.~2010a). They are characterised by differential spectral index in the range $0.7\div 1.4$, showing the maximum of the emission at a few GeV and energy of the exponential cut-offs have been estimated in the range $1.0\div 2.6$ GeV. The $\gamma$-ray power of specific GCs can differ by an order of magnitude. It seems not to be directly related to the mass of the cluster. Moreover, $\gamma$-ray spectra from GCs have very similar features to the average spectrum of MSPs in the Galactic field (see Tables in Abdo et at~2009b).

\subsubsection{TeV emission}

\begin{figure}[t]
\vskip 5.truecm
\includegraphics{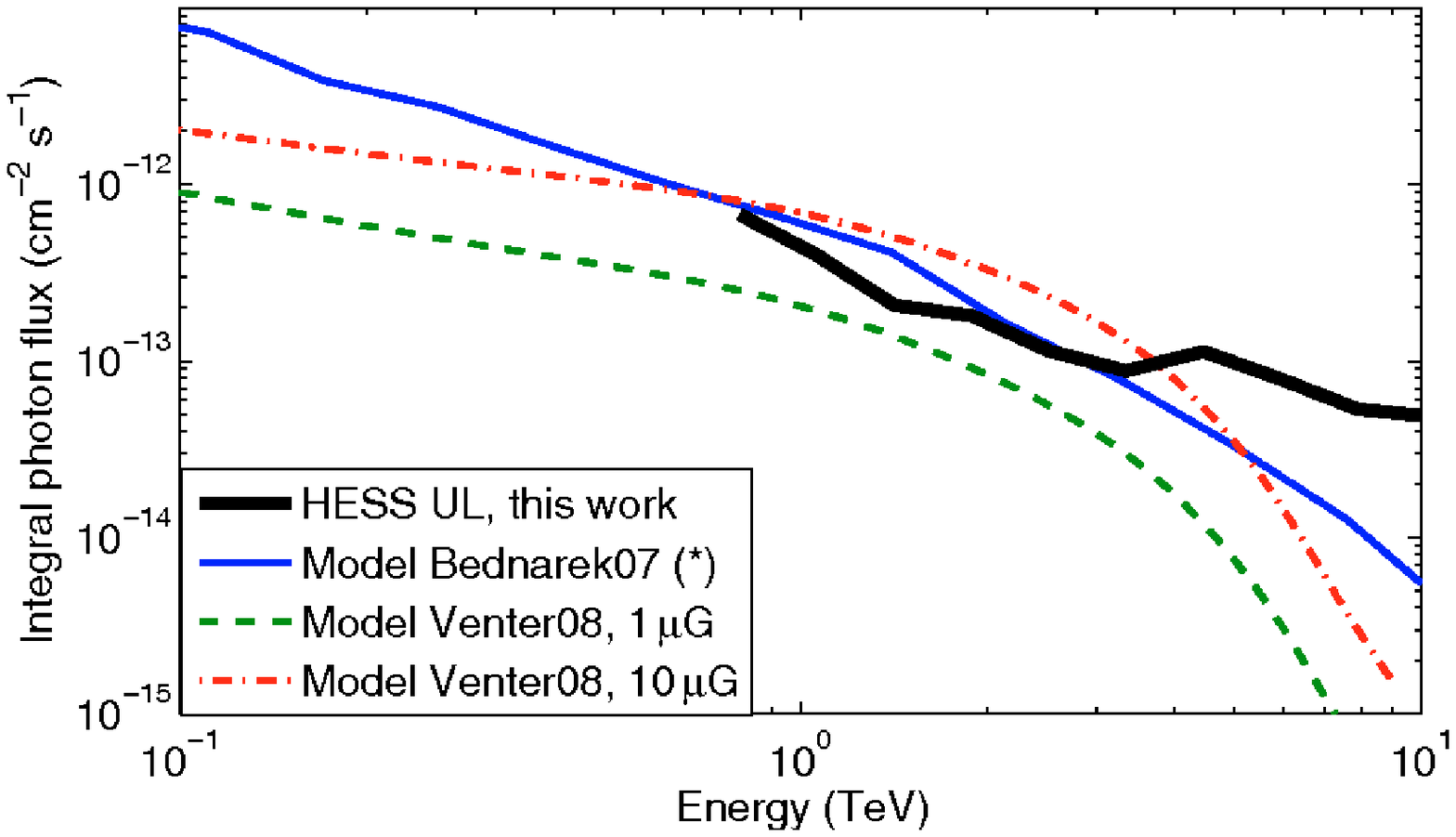}
\includegraphics{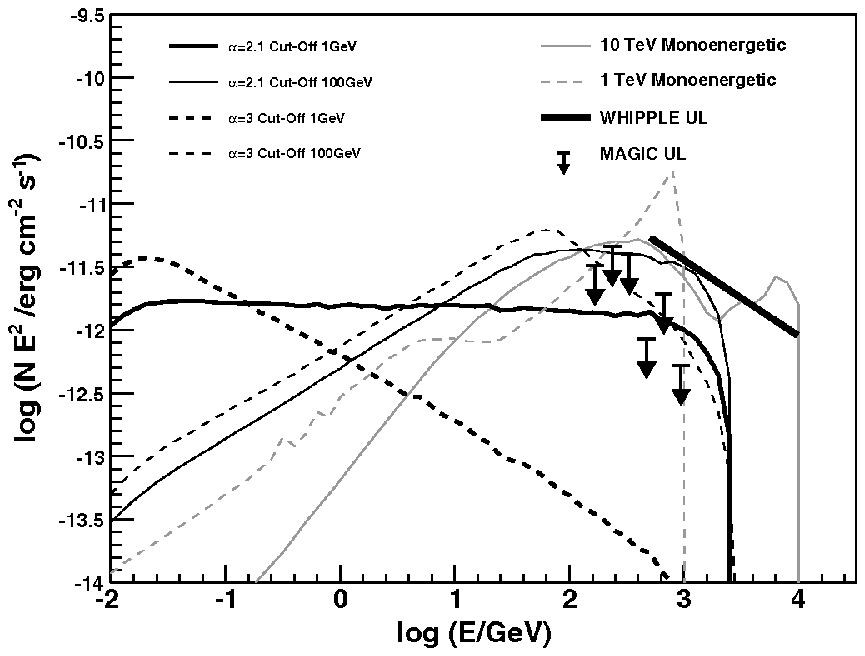}
\caption{Upper limits on the TeV $\gamma$-ray emission from two globular clusters:
Tuc 47 (left figure, thick solid curve, from Aharonian et al.~2009, reproduced by permission of the A\&A) and M 13 (right figure, arrows, from Anderhub et al.~2009, reproduced by permission of the AAS). The additional curves show the predictions of the models by Bednarek \& Sitarek~(2007) and Venter, de Jager \& Clapson~(2009).}
\label{fig2}
\end{figure}

GCs became interesting targets also for the Cherenkov telescopes at TeV energies. 
First upper limits have been reported based on the observations of two GCs, M 13 and M 15, by the Whipple Collaboration (Hall et al.~2003, Le Bohec et al.~2003). Recently, new much more sensitive instruments have also reported upper limits from directions of some
GCs. For example, the MAGIC Collaboration (Anderhub et al.~2009) observed M 13 for more than 20 hours and reported an upper limit  at energies $E_\gamma>140$ GeV, shown in Fig.~\ref{fig2}. This limit already starts to constrain the population of the MSPs within this GC based on the model developed by Bednarek \& Sitarek~(2007). The HESS Collaboration (Aharonian et al.~2009) reported the upper limit 
for Tuc 47 at energies $>800$ GeV which is on the level of predictions of the models by Bednarek \& Sitarek~(2007) and Venter et al.~(2009), see Fig.~(\ref{fig2}).
Also the VERITAS Collaboration (McCutcheon et al.~2009), observed 3 GCs, M 5, M 13 and M 15, reporting the upper limits above $600$ GeV on the level of $0.6\%$, $2.2\%$, and $1.6\%$ of the Crab Units, respectively. These last limits allow to constrain the population of MSPs within GCs on the level of 30-50, assuming that the energy conversion efficiency from the MSPs to relativistic electrons is equal to $1\%$.  Such observations should be continued with the enlarged MAGIC and HESS telescope systems. They will be able to search for the $\gamma$-ray signal clearly in the sub-TeV energies.

\section{Models for gamma-ray emission}

Due to general similarities between radiation processes occurring in the inner magnetospheres of the millisecond and classical pulsars, these first have been also suspected to be sources of pulsed $\gamma$-rays. However, the rotational energy loss rate of the MSPs is typically orders of magnitudes lower than classical pulsars. Therefore, only the closest MSPs might be potentially detected. In fact, already the EGRET detector was able to      
see a hint of a signal from one of such objects, PSR J0218+4232 (Kuiper et al.~2004).
Recently, the ${\it Fermi}$-LAT team reported detection of $\gamma$-ray emission from several MSPs in the Galactic field (Abdo et al.~2009b). Thus, confirming general theoretical predictions concerning $\gamma$-ray emission from these sources (e.g. Bulik et al.~(2000), Luo et al.~2000, Harding et al.~2005, Venter \& de Jager~2005).

Since globular clusters contain many MSPs, their cumulative $\gamma$-ray emission might be also above the sensitivity of $\gamma$-ray telescopes. A few estimates of the $\gamma$-ray 
fluxes from MSPs in globular clusters have appeared before launching the ${\it Fermi}$ Observatory. Harding et al.~(2005), based on the pair starved polar cap model, predicted 
the $\gamma$-ray fluxes from Tuc 47 near the upper limits from the EGRET. Venter \& de Jager~(2008), applying the modern version of the polar cap model (general relativistic effects included), also predicted $\gamma$-ray spectra clearly above the sensitivity of the ${\it Fermi}$-LAT telescope. 

A significant part of MSPs, detected in globular clusters, are within binary systems. MSP winds can interact with the winds of companion stars creating shocks which can accelerate electrons (e.g. Klu\'zniak et al.~1988, Phinney et al.~1988, Arons \& Tavani~1993). 
The X-rays from MSPs magnetospheres can illuminate the companion star heating it to significantly larger  temperatures than expected from pure nuclear burning (e.g. Bednarek \& Pabich~2010). Relativistic 
electrons interact with this enhanced stellar radiation producing $\gamma$-rays. Some of these MSPs can be completely surrounded by the stellar
winds creating the so called "hidden" MSPs (Tavani~1991). Such MSPs in compact binary systems
can in principle also contribute to the $\gamma$-ray emission observed from the globular clusters. 
 
Another scenario for GeV $\gamma$-ray emission can be envisaged in the case of accreting, fast rotating 
neutron stars and possibly also white dwarfs (see recent model by Bednarek~2009). In the case of slowly rotating neutron stars, the matter from the companion star can penetrate deep into the inner NS magnetosphere. It is stopped at some distance from the NS surface as a result of the interaction with rotating magnetic field. A turbulent transition region is created. Electrons can be accelerated in such a turbulent region to $E_{\rm e}>GeV$ energies. They interact with the thermal radiation from the NS surface (and/or the accretion disk) producing GeV $\gamma$-rays.

\begin{table}[t]
\vskip 5.truecm
\includegraphics{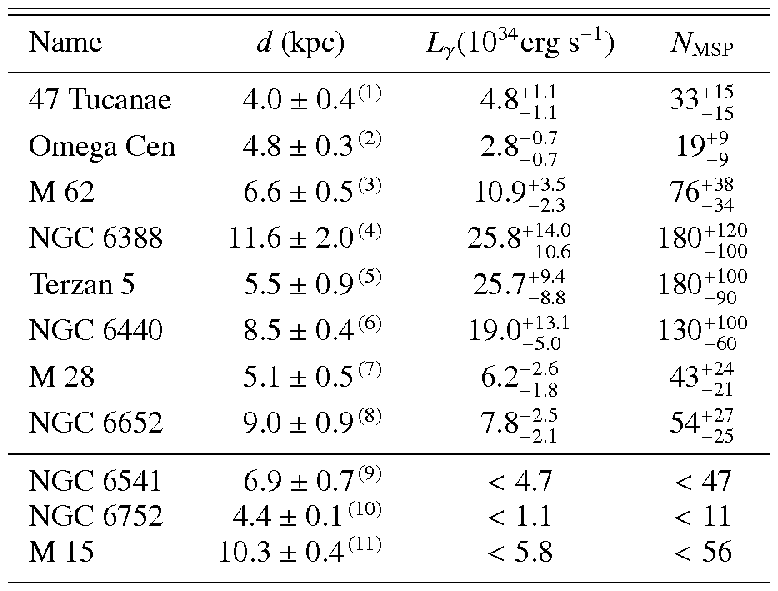}
\caption{The $\gamma$-ray luminosities from several globular clusters detected by the ${\it Fermi}$-LAT telescope. Also the expected number of millisecond pulsars estimated based on the observed $\gamma$-ray luminosity, efficiency of energy conversion from the pulsars to $\gamma$-rays and the average spin-down power of pulsars (from Abdo et al.~2010a, reproduced by permission of the AAS).}
\label{tab1}
\end{table}
\subsection{Interpretation of the observed GeV $\gamma$-ray emission}

General features of the $\gamma$-ray spectra observed by ${\it Fermi}$-LAT from GCs (spectral indices, cut-offs)
are incredibly similar to the spectra detected by this same telescope from the 
population of MSPs observed in the galactic field and from the classical pulsars (Abdo et al.~2009b). Therefore, it seems obvious that the main contribution to the GC $\gamma$-ray emission is a cumulative 
emission from the inner magnetospheres of the whole population of MSPs. 
In fact, such general $\gamma$-ray emission from MSPs fits quite well to the earlier predictions made based on the polar cap model (Harding et al.~2005, Venter \& de Jager~2008). However, there are at present some doubts concerning the validity of the polar cap model as a likely scenario for the $\gamma$-ray emission from the pulsar inner magnetospheres.
Recent observations of the pulsed $\gamma$-ray spectrum from the Crab pulsar at energies above $\sim 10$ GeV (Aliu et al.~2008, Abdo et al.~2010b) 
are clearly inconsistent with the $\gamma$-ray production region close to the stellar surface. Similar features of the $\gamma$-ray emission from the classical and the millisecond pulsars suggest that they are working under similar mechanism probably located
farther from neutron star surface but closer to the light cylinder radius.

\subsection{TeV $\gamma$-ray emission from MSP winds and shocks}

The possibility of the TeV $\gamma$-ray emission from GCs has been at first discussed by
Bednarek \& Sitarek~(2007). Here
we follow the general scenario for $\gamma$-ray production within GCs proposed in this paper. It is assumed that leptons are accelerated in the millisecond pulsar winds,  shocks created by such winds (as a result of collisions with the stellar winds or between themselves), or they are injected from the inner pulsar magnetospheres. 
These energetic leptons diffuse through the volume of the GC interact at
the same time with its soft radiation content, i.e with the optical photons from stellar population and the CMBR. We assume that the diffusion process of electrons is determined only by the magnetic field strength within GC. The simplest diffusion model is applied assuming Bohm diffusion coefficient, $D_{\rm dif} = R_{\rm L}c/3$, where $R_{\rm L}$ is the Larmor radius of leptons, and $c$ is the velocity of light.  
Due to the lack of detailed knowledge on the spectral features of leptons injected by the MSPs,
a few different models for electrons are considered with the limitations expected from the comparison with classical pulsar population.

\subsubsection{Relativistic leptons from MSPs}

The spectra of leptons injected from the MSPs into the surrounding is simply an enigma.
Recent detections of GeV $\gamma$-ray emission from MSPs in the Galactic field allow to
estimate the energy conversion efficiency from the rotating pulsar to the $\gamma$-rays.
It is on the average of the order of $\eta_\gamma\sim 0.1$ (see Abdo et al.~2009b). 
The observed $\gamma$-ray luminosities of GCs and $\eta_\gamma$ (assuming that MSPs in GCs and Galactic field are similar) allows us to estimate the total energy loss rate of MSPs in specific GC on  $L_{\rm rot}^{\rm MSP}\approx L_{\gamma}^{obs}/\eta_\gamma$.
The relativistic leptons leaving the light cylinder radius take a part, $\eta_{\rm e}$, of $L_{\rm rot}^{\rm MSP}$. This power can be also related to the magnetization parameter of the pulsar wind at the light cylinder radius of the pulsar, $\sigma_{LC}$. These coefficients have been estimated in the case of classical pulsars based on the various models, e.g.
$\sigma_{\rm LC}\sim 10^4$ (Cheng et al.~1986), $\eta_{\rm e}\sim \eta_\gamma\sim 0.1$ (Harding et al.~2002), or $\eta_{\rm e}\sim 0.01$ (Venter \& de Jager~2008).
Based on these values we can only say that the power in relativistic electrons leaving 
the light cylinder radius can be of the order, $L_{\rm e}\sim L_{\rm rot}^{\rm MSP}/\sigma_{\rm MSP}\approx L_{\gamma}^{\rm obs}/(\eta_\gamma\sigma_{\rm MSP})\sim 
(10^{-4}\div 1)\times L_{\gamma}^{\rm obs}$. These leptons can be additionally re-accelerated in the pulsar wind zone and/or the pulsar wind shock. Therefore, the above 
estimate on the power of injected leptons should be considered as the lower limit.

The power in relativistic leptons accelerated above the light cylinder radius could be in principle estimated based on the value of the magnetization parameter of the pulsar wind,
which is the ratio of the energy density of the magnetic field to the energy density of leptons in the wind. Unfortunately, we can not estimate this value for the MSP population due to the lack of observational constraints. Therefore, we base on the analysis done for the classical pulsars keeping in mind that the processes in their magnetospheres seem to be identical to those in MSPs. From modelling of two well known pulsars and their nebulae, the magnetization parameter at the pulsar wind shock has been estimated to be $\sigma << 1$ for the Crab Nebula (Kennel \& Coroniti~1984), and $\sigma\sim 0.1$ in the case of Vela pulsar (Sefako \& de Jager~2003). In the case of the pulsar wind which terminates closer, e.g. due to the interaction with a companion star, $\sigma$ could be of the order of $\sim 1$ as expected in the model by Contopoulos \& Kazanas~(2002). Such small values for the magnetization parameter at the pulsar wind or shock means that the energy conversion efficiency from the pulsar wind into the relativistic leptons is $\eta_{\rm e}\sim 1$. In such a case, the power 
in relativistic leptons injected into the GC can be estimated on $L_{\rm e}^{\rm wind}\sim 10L_{\gamma}^{obs}$, where $L_{\gamma}^{obs}$ is the pulsed $\gamma$-ray power produced in the inner pulsar magnetospheres which has been detected by the ${\it Fermi}$-LAT telescope
from specific GCs. 

The spectrum of leptons injected through the light cylinder radius could be in principle
calculated in a specific model for $\gamma$-ray production (provided that this model is 
known!). The spectrum of leptons injected from the pulsar wind shock is expected to be of
the power law type with the low energy cut-off (corresponding to energies of leptons injected from the light cylinder radius) and the high energy cut-off which could be estimated based on the known magnetic field strength at the wind shock.
We stress that up to now we have no any observational signatures from direct observations of the MSPs at very higher energies since none has been detected above  $10$ GeV. Also  detailed models for $\gamma$-ray production (and lepton injection) in the inner magnetospheres are not at present available. Therefore, the cut-offs in the injected spectrum of leptons have to be assumed rather arbitrarily. It will be simply  tested with the future observations in the GeV-TeV energy range.

Some re-scaling from observations of the pulsar wind nebulae around classical pulsars in the TeV $\gamma$-rays (e.g. the Crab nebula) gives the values for the maximum energies of leptons from MSP winds of the order of, 

\begin{eqnarray}
E^{\rm max}_{e}\sim 3\times 10^{15}\star ({{3\times 10^8 G}\over{4\times 10^{12} G}}) \star ({{4 ms}\over{33 ms}})^{-2} eV\sim 15 TeV.
\label{eq3}
\end{eqnarray}

Specific models for TeV $\gamma$-ray production in GCs apply the values of similar order.
For example, based on the comparison of the advection time of leptons along the shock with the acceleration time, Bednarek \& Sitarek~(2007) apply $E_{\rm e}^{\rm max}$ in the range  4-40 TeV. On the other hand, Venter et al.~(2009) has got similar values by calculating the injection spectrum of leptons through the light cylinder radius by integrating over the whole population of MSPs in specific GC applying the polar cap model.

\subsubsection{Gamma-ray spectra from leptons diffusing inside GC}

Leptons with spectra discussed above are injected into the radiation field of the
GC and diffuse in the outward direction comptonizing the soft radiation field within the cluster (stellar, CMBR, infrared).

\begin{figure}[t]
\vskip 11.7truecm
\includegraphics{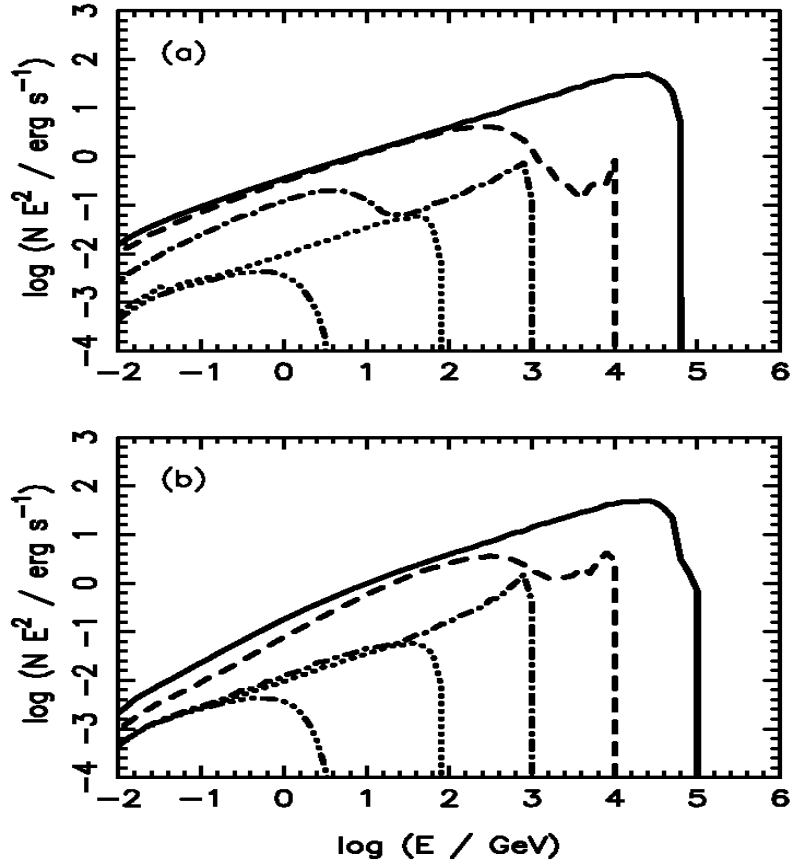}
\caption{Differential $\gamma$-ray spectra (multiplied by energy squared) produced in the IC scattering of the stellar and CMBR radiation by mono-energetic leptons with energies:
10 GeV (triple dot-dashed curves, $10^2$ GeV (dotted), $10^3$ GeV (dot-dashed), and $10^4$ GeV (solid). Leptons are injected from MSPs and diffuse in the GC magnetic field with the strength $10^{-6}$ G. The total stellar luminosity of the cluster is $10^5L_\odot$ (a) 
and $10^6L_\odot$ (b). The spectra are normalized to single lepton injected in 1 second(from Bednarek \& Sitarek~2007).}
\label{fig3}
\end{figure}

The basic proprieties of the $\gamma$-ray spectra in the case of mono-energetic injection of leptons with different energies into the soft radiation field dominated by soft photons
from the stellar population within the cluster and the MBR are shown in Fig.~\ref{fig3} (Bednarek \& Sitarek~2007).  GCs with two different stellar luminosities are considered. Diffusion process of leptons is followed up to 100 pc from the center of GC and an average magnetic field strength to be of the order of $B_{\rm  GC}-10^{-6}$ G. 
Note the interesting features of these $\gamma$-ray spectra. For leptons with TeV energies two features in the $\gamma$-ray spectrum appear to be caused by the scattering of the MBR in the Thomson regime and the scattering of optical photons in the Klein-Nishina (KN) regime. Interestingly, the lower energy feature resambles very much a bump of GeV $\gamma$-ray emission reported by the ${\it Fermi}$-LAT from the globular clusters. The second feature, a sharp KN peak, has not been observed up to now. This is not so surprising since it is predicted in the sub-TeV energy range marginally accessible by the present Cherenkov telescopes.

\begin{figure}[t]
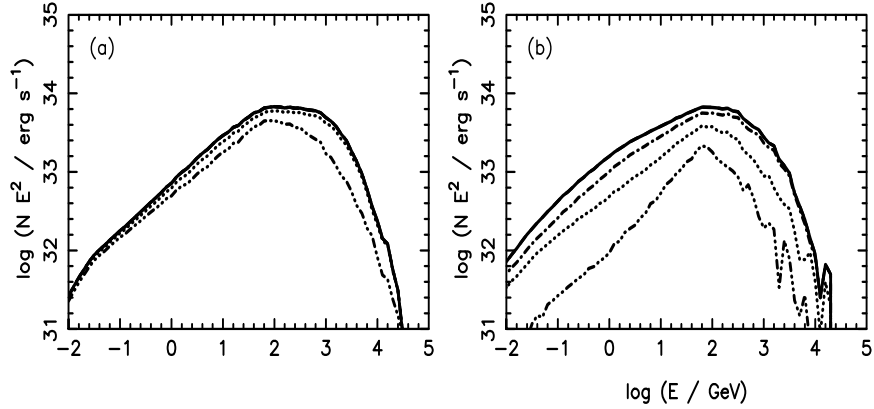

\vskip 5.8truecm
\includegraphics{bednarf4a.eps}
\includegraphics{bednarf4b.eps}
\caption{As in Fig.~3 but for the power law spectrum of leptons with the spectral index 2.1 between 100 GeV and 30 TeV. Specific curves show the spectra within the central region of GC
equal to 1 pc (triple-dot curve), 3.3 pc (dotted), 10 pc (dot-dashed), 33 pc (dashed), and 100 pc (solid). The magnetic field within the cluster has been fixed on $10^{-5}$ G (a) and $10^{-6}$ G (b). The spectrum of leptons is normalized to the 30 MSPs with surface magnetic field $10^9$ G and period 4 ms within the cluster at the distance of 5 kpc. (from Bednarek \& Sitarek~2007).}
\label{fig4}
\end{figure}

The example $\gamma$-ray spectra produced by leptons which are injected with the power law spectrum at specific energy range are shown in Fig.~\ref{fig4}. 
These spectra show a maximum at $\sim$0.1-1 TeV which is produced by the
scattering of soft radiation in the Thomson regime. At higher energies the spectral index of $\gamma$-ray emission changes due to the scattering in the KN regime.  
The $\gamma$-ray spectra produced within specific volume around the center of GC are also shown on this figure. For stronger magnetic fields within GC
($B_{\rm GC} = 10^{-5}$ G), most of the emission is confined within the volume of GC, i.e. it creates a point like source for Cherenkov telescopes.

\begin{figure}[t]
\vskip 9.2truecm
\includegraphics{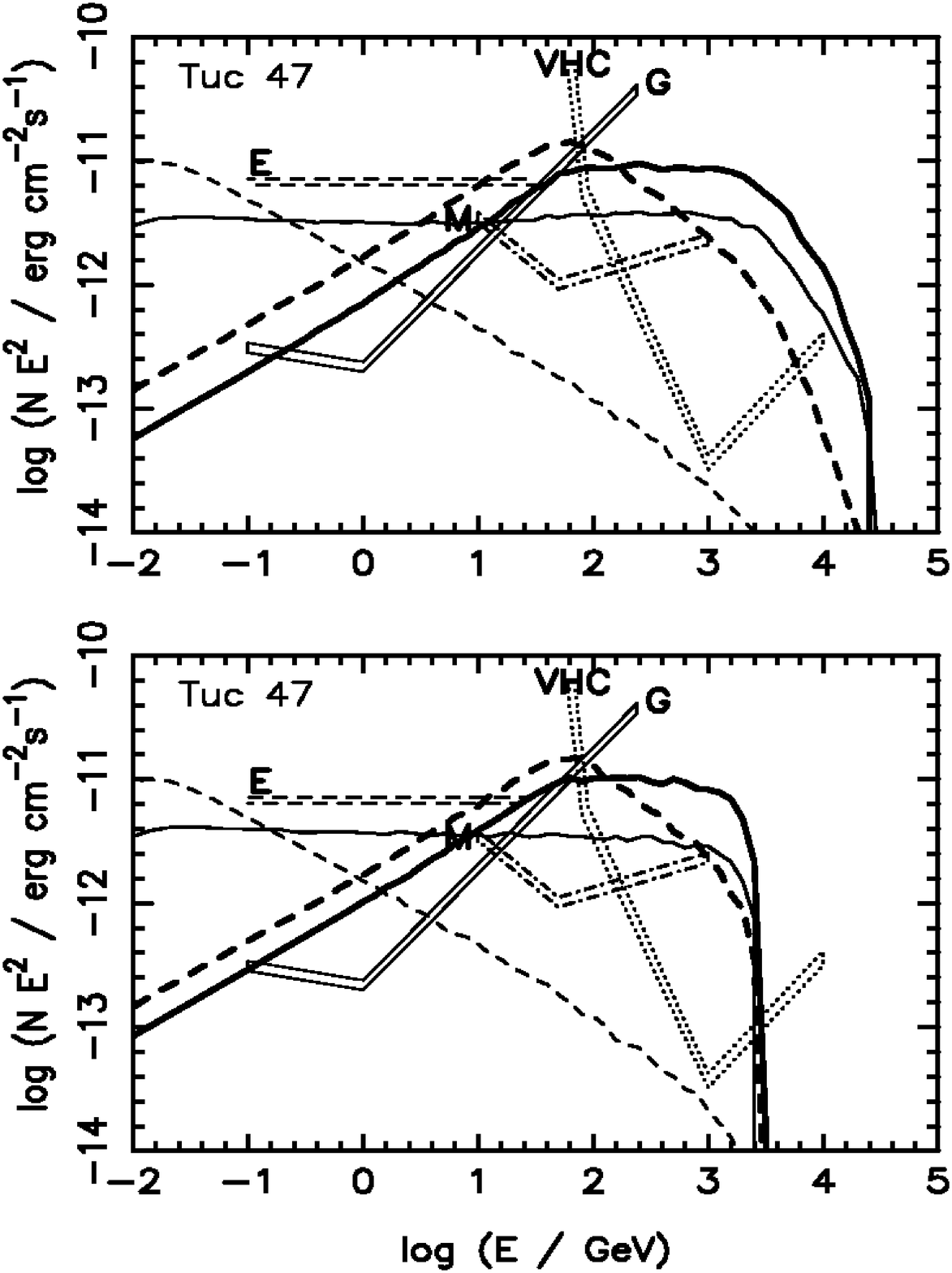}
\includegraphics{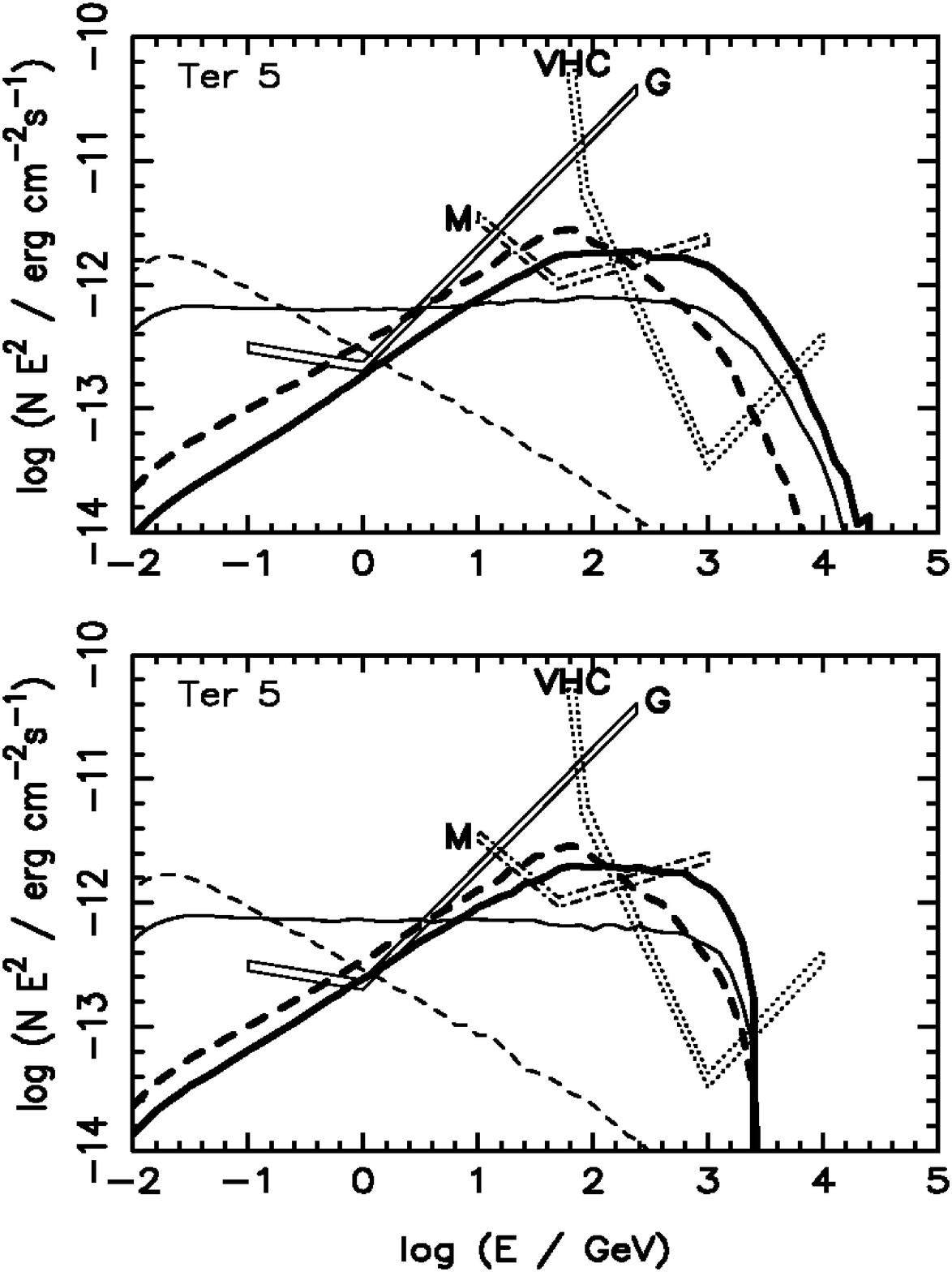}
\caption{Comparison of the $\gamma$-ray spectra expected from Tuc~47 (left) figures) and Ter~5 (right) with the sensitivities of different instruments (E - EGRET, G - ${\it Fermi}$-LAT, V -Veritas, H - HESS, and M - MAGIC).
The sensitivities of Cherenkov telescopes are for the 5$\sigma$ detection within 50 hrs. The $\gamma$-ray spectra has been calculated for different injection spectra of leptons from MSPs (from Bednarek \& Sitarek~2007). }
\label{fig5}
\end{figure}

In terms of such a model, the $\gamma$-ray spectra have been calculated for selected globular clusters assuming different spectra of injected leptons (see Fig.~\ref{fig5}, Bednarek \& Sitarek~2007).
This emission depends on the total power in leptons injected from MSPs, which is determined by the number of MSPs in specific GC and the efficiency of lepton acceleration by the average pulsar. Therefore, the comparison of the predicted spectra with the observations of specific GCs allows us to put the upper limits on the product of these two free parameters. Such procedure has been performed by Bednarek \& Sitarek~(2007), applying the sensitivities of present Cherenkov telescopes.

\begin{figure}[t]
\vskip 9.truecm
\includegraphics{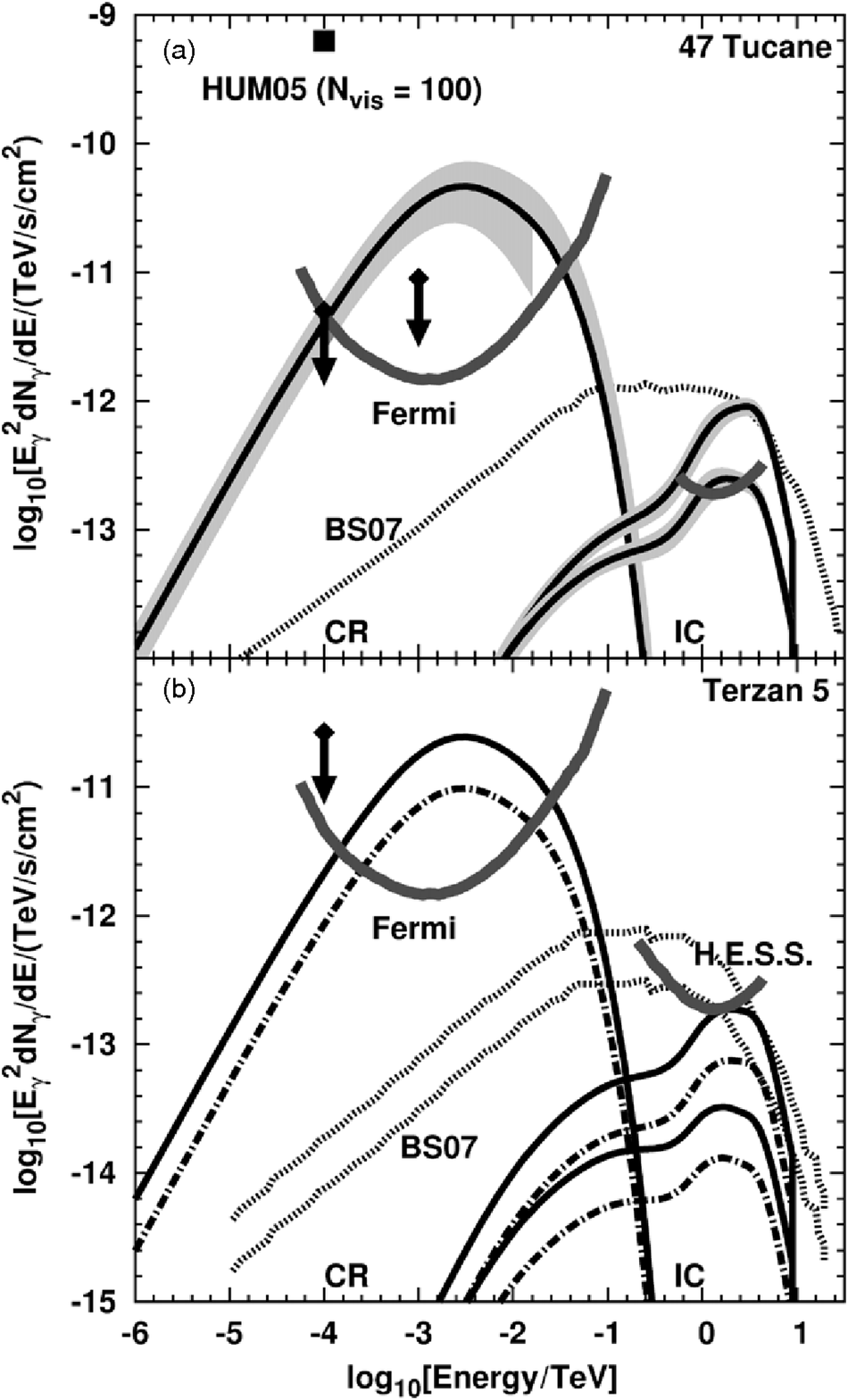}
\caption{Differential $\gamma$-ray spectra produced in the MSPs inner magnetospheres (curvature) and in the MSPs surrounding (inverse Compton) expected in the polar cup model
for two globular clusters Tuc 47 and Ter 5. The magnetic field in Tuc 47 is assumed 10$\mu$G
and in Ter 5 1$\mu$G. The $\gamma$-ray flux predicted by Harding et al.~(2005) is marked by HUM05 and by Bednarek \& Sitarek~(2007) by BS07.
The ${\it Fermi}$-LAT and HESS sensitivities and the EGRET upper limits are also marked  
(from Venter, de Jager \& Clapson~2009, reproduced by permission of the AAS).}
\label{fig6}
\end{figure}

Similar model has been considered more recently by Venter, de Jager \& Clapson~(2009). In this calculations it is assumed that leptons are injected from the inner millisecond pulsar magnetospheres through the light cylinder radii in terms of an isolated pulsar polar cap model (Venter \& de Jager~2008). In such a case, the injection rate of leptons is directly linked to the pulsed $\gamma$-ray emission from the MSPs observed in the GeV energies.
The results of calculations are shown in Fig.~\ref{fig6} for two GCs, Tuc 47 and Ter 5.
The TeV $\gamma$-ray emission expected in this work is quite similar to those predicted by Bednarek \& Sitarek~(2007). Some differences are predicted at the sub-TeV $\gamma$-rays
due to different shapes of assumed spectra for leptons. 
However, note that the GeV $\gamma$-ray emission predicted by Venter et al.~(2009) for applied example parameters of the MSP population overestimates the EGRET upper limit for Tuc 47 (and also ${\it Fermi}$-LAT detection) by a factor of a few. Due to the lower $\gamma$-ray flux reported by the ${\it Fermi}$-LAT telescope from Tuc 47, the TeV $\gamma$-ray emission predicted by Venter et al.~(2009) should be reduced by this same factor being on the level below sensitivity of the present HESS telescope system.

\subsection{Constraints on the MSP population}

The GCs identified with the LAT $\gamma$-ray sources differ in GeV $\gamma$-ray power by an order of magnitude (Abdo et al.~2010a). This power seems to be correlated with the stellar encounter rate, derived for a specific GC. It is expected that the encounter rate determines the formation rate of compact binaries which are also responsible for the origin of the MSPs inside GCs. The encounter rate, $\Gamma_{\rm e}$, can be estimated from the observed density of stars in the core of GC and the core radius (Gendre et al.~2003). Based on the observed $\gamma$-ray luminosity of specific GC, the average energy loss rate of MSP (adopted for MSPs observed in Tuc 47, Abdo et al.~2009a), and the average spin-down to $\gamma$-ray luminosity conversion efficiency (Abdo et al.~2009a), Abdo et al.~(2010a) estimate the number of MSPs, $N_{\rm MSP}$, in specific GC. This number is confronted with the encounter rate (see Fig.~3 in Abdo et al.~2010a). The correlation between these values is well fitted by $N_{\rm MSP} = (0.5\pm 0.2)\times \Gamma_{\rm e} + (18\pm 9)$. 
Note that, the populations of MSPs in specific GCs derived by Abdo et al.~(2010a) are
in good agreement with the estimates of the number of MSPs obtained with other methods (see e.g. estimates for Ter 5 by Fruchter \& Goss~(2000) and Kong et al.~(2010)).

\begin{figure}[t]
\vskip 7.3truecm
\includegraphics{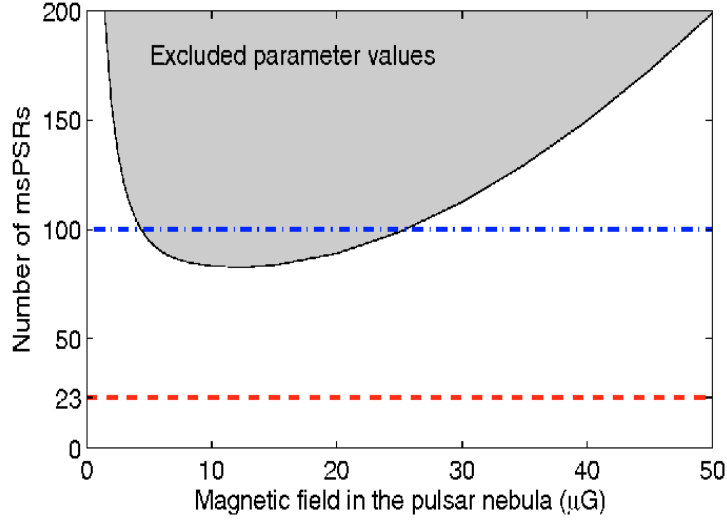}
\caption{Upper limit on the number of MSPs in Tuc 47 as a function of the magnetic field strength within the pulsar nebula based on the model of Venter et al.~(2009) and the HESS flux upper limit (from Aharonian et al.~2009, reproduced by permission of the A\&A).}
\label{fig7}
\end{figure}

The theoretical estimates of the TeV $\gamma$-ray fluxes triggered recent
observations of some GCs by Cherenkov telescopes (Aharonian et al. 2009, Anderhub et al.~2009, McCutcheon et al.~2009). Unfortunately, only upper limits have been reported up to now. Based on the comparison of the models with these upper limits, it is possible to put constraints on the total energy in relativistic leptons which depend on the number of the MSPs within specific GC and the efficiency of energy conversion from rotating MSP to relativistic leptons. The results for a few northern hemisphere
GCs (M13, M15, M5) are reported by the MAGIC and VERITAS Collaborations (Anderhub et al.~2009, McCuteon et al.~2009). The limits obtained for these GCs are still consistent with the expected population of MSPs in specific objects.
The HESS Collaboration (Aharonian et al.~2009) reported upper limits for Tuc 47. Based on these limits the parameter space defined by the average strength of the magnetic field within GC and the number of millisecond pulsars has been constrained
(see Fig.~\ref{fig7}). In the most limiting case, the HESS upper limits constrain the number of MSPs in Tuc 47 on $\sim 80$.
All these above mentioned upper limits are not able to exclude the possibility of injection of leptons with the TeV energies into the GC radiation field. Future, more sensitive observations with the stage II Cherenkov arrays (HESS II, MAGIC II, new VERITAS arrangement), are very welcomed in order to start to constrain the models for acceleration of leptons by MSPs.

\section{Gamma-rays from electrons injected from other sources ?} 

We wonder whether particles injected from other sources or within other scenarios for the present radiative processes in the MSP surroundings might also contribute to the $\gamma$-ray flux observed by the ${\it Fermi}$-LAT telescope on the position of detected GCs.
Let us consider the injection of electrons by other sources than MSPs within GCs. 
Note that, the spectra of leptons injected by the MSPs can be also characterised
by different components as seems to be clear from modern observations and modelling of classical pulsars (e.g. polar and equatorial components, dependence of the spectrum of injected leptons on the angle above/below the rotational plane of the pulsar). 
We assume that electrons are injected with some specific energies from explicitly undefined source. These electrons might contribute to GeV $\gamma$-ray emission producing also some characteristic features in the multiwavelength spectrum which might be tested by the observations at other parts of the electromagnetic spectrum.

\begin{figure}[t]
\vskip 6.1truecm
\includegraphics{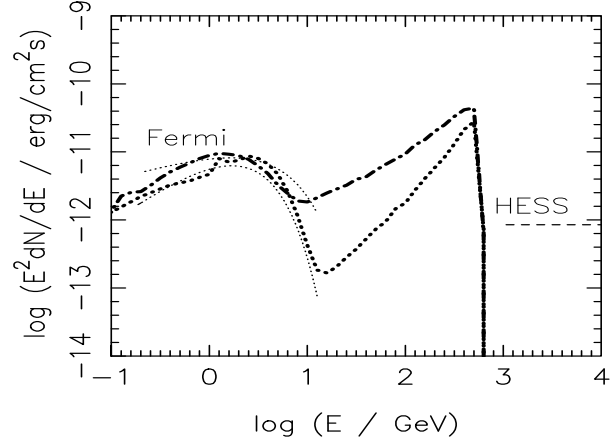}
\caption{The example $\gamma$-ray spectra produced by electrons in the ICS of the MBR and stellar radiation. Electrons with energies $E_{\rm e}\sim 600$ GeV diffuse in the magnetic field $10^{-6}$ G through the globular cluster Tuc 47. 
Diffusion process of electrons occurs on a time scale $\tau_{\rm dif} = 0.01\tau_{\rm Bohm}$ (dot-dashed curve) and $0.001\tau_{\rm Bohm}$ (dotted). The lower energy bump is produced by scattering of the MBR and the sharp higher energy bump by scattering stellar photons.}
\label{fig8}
\end{figure}

We investigate the hypothesis that characteristic bumps observed in the GeV $\gamma$-rays are produced as a result of the Inverse Compton Scattering (ICS) of the CMBR by electrons with typical energies of the order of $E_{\rm e}\sim 600$ GeV. Such electrons should mainly contribute to the GeV energy range (showing also characteristic cut-off at a few GeV )as a result of scattering of CMBR (estimated from $E_\gamma\approx \varepsilon_{\rm MBR} (E_{\rm e}/m_{\rm e})^2$). Note that electrons with such energies should also produce synchrotron radiation. For the magnetic field of the order of expected within GCs,
$B_{\rm GC}\sim 10^{-6}-10^{-5}$ G, typical energies of synchrotron photons
fall at energies, $E_{syn}\sim m_e(B/B_{cr})\gamma_e^{2}\sim 0.05-0.5$ eV, i.e. in the optical range. The power emitted in these optical photons should be within an order of magnitude of the power observed in GeV $\gamma$-rays. Such emission can not be observable due to much stronger radiation field coming from the stellar population in the globular cluster. On the other hand, such relativistic electrons should comptonize stellar radiation in the KN regime producing $\gamma$-rays with the spectrum extending up to $\sim 600$ GeV. The power in this radiation depends on the density of stellar photons (which can be even comparable to the density of the microwave background photons) and the diffusion time scale of electrons through the volume of the globular cluster. In order not to over-produce sub-TeV $\gamma$-ray emission in such a scenario, the diffusion time scale of electrons has to be much shorter than the Bohm diffusion.
We have performed example calculations of the $\gamma$-ray spectra in terms of such scenario for the globular cluster Tuc 47, assuming that diffusion of electrons occurs on the time scale equal to: $\tau_{\rm dif} = 0.01\tau_{\rm Bohm}$ 
and $0.001\tau_{\rm Bohm}$ (see Fig.~\ref{fig8}). With these parameters, the GeV $\gamma$-ray emission measured from Tuc 47 can be reasonably described by the ICS of the MBR. The hard $\gamma$-ray spectrum, which extends above $\sim 10$ GeV and cuts-off abruptly at a few hundred GeV, can not be constrained by the present upper limits derived by the HESS Cherenkov array. Note that the $\gamma$-ray emission envisaged in such a scenario should 
have angular extend larger than the typical size of the globular cluster.
This extend might be studied with the present and future $\gamma$-ray telescopes and serve as a test of the possible contribution of a few hundred GeV electrons to the GeV $\gamma$-ray bump.

\begin{figure}[t]
\vskip 4.7truecm
\includegraphics{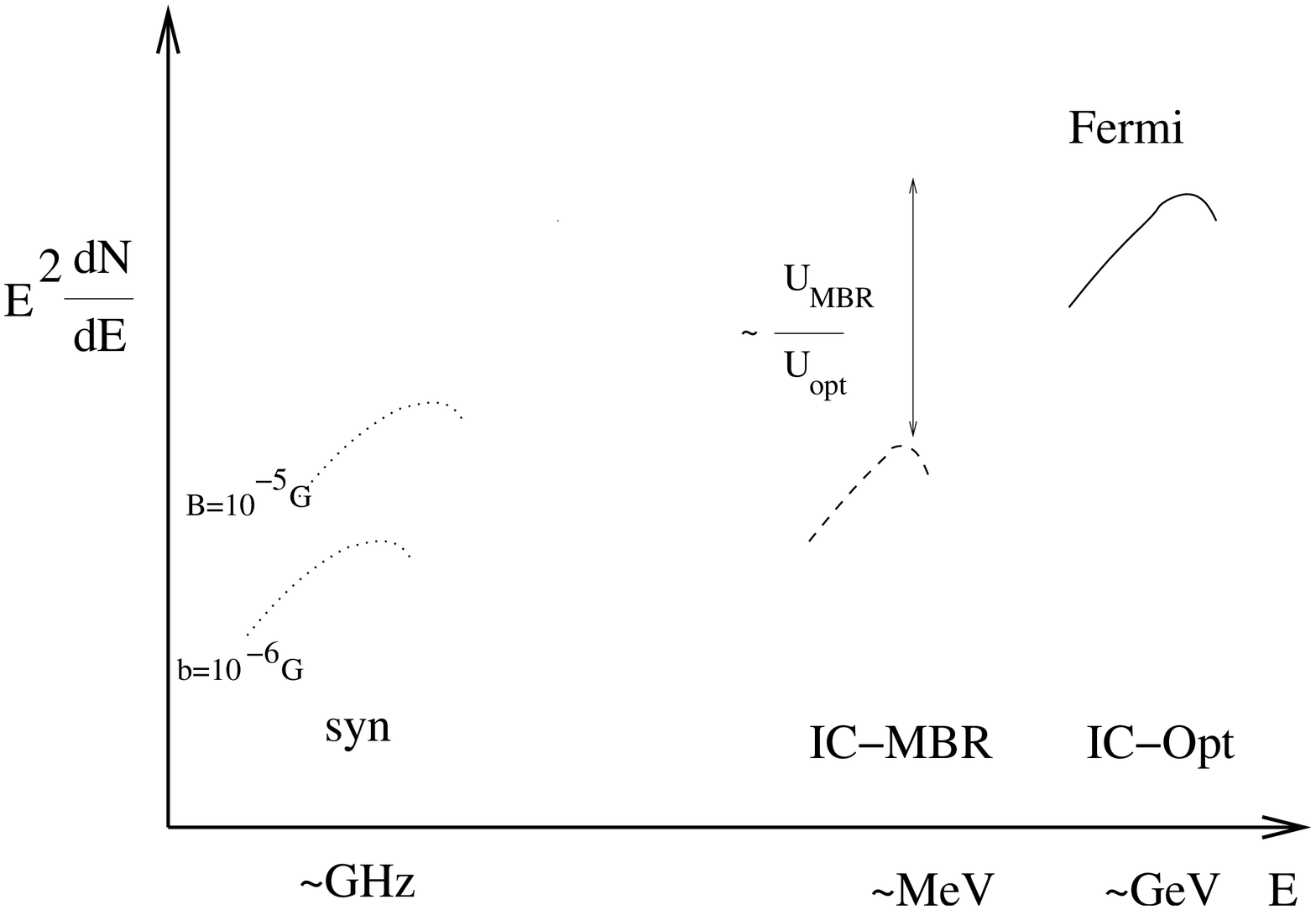}
\caption{As in Fig.~8 but for electrons with energies $E_{\rm e}\sim 20$ GeV which diffuse in the magnetic field of the order of $10^{-6}-10^{-5}$ G through the globular cluster Tuc 47. Diffusion process of electrons occurs on the Bohm diffusion time scale. The ${\it Fermi}$-LAT $\gamma$-ray spectrum is produced by ICS of stellar radiation, the soft $\gamma$-rays are produced by scattering MBR and the synchrotron radio emission is due to synchrotron process.}
\label{fig9}
\end{figure}

In principle, the characteristic $\gamma$-ray spectrum measured by {\it Fermi}-LAT can be also explained as a result of ICS of stellar photons by electrons with typical energies of the order of $\sim 20$ GeV. In fact, the IC cooling time scale for electrons with such energies is comparable to the Bohm diffusion time scale of such electrons within the globular cluster, $\tau_{\rm IC}^{\rm opt}\sim \tau_{\rm dif}^{\rm Bohm}\sim 1.5\times 10^4$ yrs. Therefore, it is expected that energy is transferred efficiently from electrons to GeV $\gamma$-rays. Moreover, these electrons will also produce soft $\gamma$-rays as a result of comptonization of the MBR
with typical energies, $E_\gamma^{\rm MBR}\sim \varepsilon_{\rm opt}\gamma_{\rm e}^2\sim MeV$. Since the IC scattering of optical photons and the MBR occurs in the Thomson regime, the relative power in the MeV $\gamma$-rays should be on the level of $\sim U_{\rm MBR}/U_{\rm opt}\sim 10^{-3}-10^{-2}$. The MeV $\gamma$-ray emission on such a level is at present undetectable by any instruments. Since the magnetic field energy density within the globular cluster is expected to be of the order of the energy density of the MBR then, we expect that synchrotron emission with similar power to that one produced in ICS process of the MBR should be also expected. Typical energies of these synchrotron photons are 
$\varepsilon_{\rm syn}\approx m_{\rm e}c^2(B/B_{\rm cr})\gamma_{\rm e}^2$, i.e. they fall into $\sim$GHz frequencies. Radio emission with such power should be detectable by the present telescopes. Therefore, discovery of such diffuse radio emission from the globular clusters can provide a test for the mechanism discussed above. The general features of the broad band emission expected in this last possibility are shown in Fig.~\ref{fig9}.

In summary, other mechanisms or sources, than magnetospheric emission from 
the population of the millisecond pulsars, may also contribute to the characteristic bump in the $\gamma$-ray spectrum detected by {\it Fermi}-LAT from several globular clusters.

\section{Conclusions}

Globular Clusters are abundant with different types of compact objects such as, millisecond pulsars, accreting neutron stars (LMXBs) and white dwarfs (Cataclysmic Variables) and possibly also black holes. These compact objects are expected to accelerate particles. They find well defined radiation field from whole population of stars. Therefore, Globular Clusters are likely sites for non-thermal high energy emission. 
In fact, it has been predicted that the population of MSPs within GCs can produce
observable fluxes of GeV $\gamma$-rays which results from radiation processes in the inner pulsar magnetospheres (Harding et al.~2005, Venter \& de Jager~2008).
The steady GeV $\gamma$-ray emission has been recently detected from several GCs by the ${\it Fermi}$-LAT telescope (Abdo et al.~2009a, Abdo et al.~2010a). This emission can be naturally explained as a cumulative contribution from the whole population of MSPs within specific GC due to clear similarities of their spectral features to the population of MSPs in the Galactic field detected in $\gamma$-rays (Abdo et al.~2009b). A single MSP has not enough power to provide observed $\gamma$-ray fluxes from GCs.
The application of the energy conversion efficiency from MSPs to $\gamma$-rays, derived for MSPs in the Galactic field, to the $\gamma$-ray emission from GCs allows to estimate 
the number of MSPs in specific GC. These numbers are consistent with the estimates
of the MSP population in these GCs based on other methods. Note however, that such method assumes that MSPs in GCs and Galactic field have similar proprieties which is questioned in some works (e.g. Hui et al.~2010). Moreover, such comparison assumes that MSPs are the only sources contributing to the GeV $\gamma$-ray emission observed from GCs. 

Other objects within GCs, might also contribute to the $\gamma$-rays at GeV energies (e.g. accreting neutron stars or white dwarfs), either in their immediate surrounding or by injecting electrons with specific energies into the volume of GCs. In fact, the characteristic features of the GeV emission (flat spectrum and the cut-off above a few GeV) can be also well described in the scenario in which electrons with specific energies scatter predominantly stellar radiation or the MBR. In both cases, characteristic emission features are predicted at other parts of electromagnetic spectrum which might serve as a diagnostic of such possible processes.  Also sources at the background field (e.g. galaxy clusters) or shocks created by GCs interacting with Galactic gas might contribute to $\gamma$-ray emission as often considered in the case of the non-thermal X-ray emission observed from GCs.

MSPs within GCs should likely accelerate leptons up to very high energies at the inner pulsar magnetospheres, pulsar wind zones and/or wind shocks. These leptons gradually diffuse in the GC magnetic field interacting with its soft
radiation field (Bednarek \& Sitarek~2007, Venter et al.~2009). As a result, TeV $\gamma$-ray emission is expected from the inverse Compton scattering of soft radiation field (stellar, MBR, and from Galactic disk). Up to now, observations of GCs with the Cherenkov telescopes (Anderhub et al.~2009, Aharonian et al.~2009, McCutcheon et al.~2009) have provided only the upper limits which already start to constrain the population of the MSPs within GCs and the conditions for the diffusing process within GCs. Note that detection of TeV $\gamma$-ray emission from the whole population of MSPs in specific GCs should provide very useful additional constraints on the acceleration and injection process of leptons from the inner pulsar magnetosphere's and pulsar winds. Therefore, they are very welcomed by the theoreticians working on pulsar electrodynamics. 
It is hoping that with the future sensitive observations with the advanced HESS, MAGIC and VERITAS telescope systems (or planned CTA) such positive detections will be possible. 

\begin{acknowledgement}
This work is supported by the Polish MNiSzW grant N N203 390834. 

\end{acknowledgement}

\end{document}